\documentclass{IEEEtran}
\usepackage{cite}
\usepackage{amsmath,amssymb,amsfonts}
\usepackage{graphicx}
\usepackage{textcomp,nicefrac}
\usepackage[colorlinks=true, linkcolor=blue, citecolor=blue, urlcolor=blue]{hyperref}
\usepackage{color,soul}

\usepackage{placeins}   
\usepackage{float}         
\usepackage{dblfloatfix}   

\begin{document}
\title{Characterization of precipitation-induced radon progeny deposition events using a city-scale sensor network}
\author{
    S. Das\IEEEauthorrefmark{1},
    N. Abgrall\IEEEauthorrefmark{2},
    M.S. Bandstra\IEEEauthorrefmark{2},
    R.J. Cooper\IEEEauthorrefmark{2},
    Y. Kim\IEEEauthorrefmark{3},
    E. Rofors\IEEEauthorrefmark{2},
    R. Sankaran\IEEEauthorrefmark{3},
    S. Shahkarami\IEEEauthorrefmark{3}\\
    \IEEEauthorblockA{\IEEEauthorrefmark{1}Department of Physics, Williams College, Williamstown, MA, USA}\\
    \IEEEauthorblockA{\IEEEauthorrefmark{2}Lawrence Berkeley National Laboratory, Berkeley, CA, USA}\\
    \IEEEauthorblockA{\IEEEauthorrefmark{3}Argonne National Laboratory, Lemont, IL, USA}%
    \thanks{This work was performed under the auspices of the U.S. Department of Energy by Lawrence Berkeley National Laboratory (LBNL) under Contract DE-AC02-05CH11231. The project was funded by the U.S. Department of Energy, National Nuclear Security Administration, Office of Defense Nuclear Nonproliferation Research and Development. E-mail: RJcooper@lbl.gov}
}
\maketitle

\begin{abstract}
Networks of radiation detectors provide a platform for real-time radioactive source detection and identification in urban environments. Detection algorithms in these systems must adapt to naturally-occurring changes in background, which requires well-characterized relationships between precipitation events and their corresponding radiological signature. We present a quantitative and qualitative description of rain-induced radon progeny deposition events occurring in Chicago from September 2023 through February 2024. We measure ambient gamma radiation levels, precipitation rate, temperature, pressure and relative humidity in a network of sensor nodes. For each identified precipitation period, we decompose spectra into static- and radon-associated components as defined by a non-negative matrix factorization (NMF) algorithm. We find a consistent power-law relationship between a precipitation-dependent peak of the radon progeny proxy (RPP) and the peak strength of radon-associated NMF component for most precipitation events. We conduct a case study of a rainfall period with abnormally high levels of implied radon progeny concentration and describe its temporal and spatial evolution. We hypothesize this phenomenon is due to the air mass path that intersects a uranium-rich region of Wyoming. Finally, we cluster precipitation events into three distinct categories. One category roughly corresponds to events with deep low-pressure systems and high relative radon concentration, while another is characteristic of light stratiform rain with slightly higher temperatures and intermediate relative radon concentration. The third category appears to contain weak-gradient or lake breeze convection showers with intermittent precipitation and low relative radon concentration. These findings suggest that radiological anomaly detection could be improved by training unique background models corresponding to each category of meteorological event. 

\end{abstract}


\section{Introduction}
\label{sec:introduction}
\IEEEPARstart{M}{odern} nuclear security efforts include the detection of anomalous radiological signatures in urban environments with high sensitivity and specificity. Networks of stationary radiation monitors at roads and junctions can help assess the position and trajectory of mobile radiation sources, which is crucial for identifying vehicles carrying potentially illicit radio-nuclear material~\cite{Brennan2004, Runkle2005, Vilim2009, Wu2019}. Integrated detection systems may incorporate video footage and lidar scans with measured gamma-ray spectra to profile traffic. The SIGMA~\cite{SIGMA} and NOVArray~\cite{NOVArray} projects have demonstrated the suitability of spatially distributed arrays of radiation detectors in urban environments. These efforts operated the detectors independently, without the ability to communicate with one another.

A key challenge in tracking anomalous radiation signals is identifying such signals among expected background variations due to the presence of nuisance sources (\textit{e.g.}, medical, industrial sources) and naturally changing environmental conditions. While in the former case, a detailed study of patterns of life and contextual information can help suppress false alarms, the latter requires understanding of how environmental conditions affect the detected signal and how to take them into account. In particular, several studies have recognized precipitation-induced $^{222}\text{Rn}$ progeny deposition as a major source of interference for projects attempting to monitor radiation levels near nuclear facilities~\cite{Bottardi2020, Burnett2010, Segovia1990}. Radon-222 is a noble gas produced in the decay chain of $^{238}$U. Uranium-238 is found in soil, concrete, brick, stone, and other common construction materials. As $^{222}$Rn is produced, it travels through capillaries within the medium until it is exhaled and transported through the atmosphere via convection, turbulent diffusion, and wind motion ~\cite{Wilkening1981}. Due to its chemical inertness and half-life of 3.84 days, $^{222}$Rn is an excellent atmospheric tracer for tracking the movement of air masses ~\cite{Arnold2010, Barbosa2017, Bottardi2020}. Radon-222 decays into a variety of short-lived metallic radioisotopes that ionize and attach to suspended atmospheric dust particles. These particles serve as nucleation sites for the ice crystals and water droplets composing clouds ~\cite{Burnett2010}. When it rains, the precipitation of raindrops or snowflakes containing scavenged radon progeny (`rainout') causes a sharp increase in surface-level gamma radiation, predominantly due to $^{214}$Pb ($t_{1/2}=27.06$ min) and $^{214}$Bi ($t_{1/2}=19.21$ min). Some below-cloud radioisotopes enter raindrops through processes such as mechanical impaction, interception, and thermophoresis~\cite{Takeuchi1982}. These contribute to a `washout' factor that is insignificant relative to the in-cloud contribution~\cite{Fujinami1995, Mercier2009}. 

Lawrence Berkeley National Laboratory’s Platforms and Algorithms for Networked Detection and Analysis (PANDA) and Argonne National Laboratory’s Domain Aware Waggle Network (DAWN) projects developed and deployed a multi-sensor, city-scale testbed throughout Chicago to study networked radiation detection~\cite{Cooper2023}. PANDA develops adaptive radiological detection methods~\cite{Bandstra2023} and network analysis algorithms. Their simulations show that communication between nodes and contextual information (\textit{e.g.}, vehicle make and color) significantly improves anomaly detection capabilities~\cite{Rofors2024}. A connected network of detector nodes benefits from the ability to fuse data from multiple below-threshold encounters into significant network-wide alarms.

Characterizing the impact of precipitation-induced radon progeny deposition is essential for producing background models resistant to environmental time-dependent fluctuations in gamma radiation. The relationship between the precipitation magnitude and gamma-ray count rate has been thoroughly explored by numerous authors~\cite{Barbosa2017, Bottardi2020, Burnett2010, Fujinami1995, Fujitaka1992, Greenfield2003, Inomata2007, Kataoka1982, Livesay2014, Mercier2009, Melintescu2018, Takeuchi1982, Yakoleva2016, Yoshioka1992}. In this work, we present a related characterization of the environmental conditions associated with precipitation-induced radon progeny deposition events, focusing on a unique event that does not fit into the established trends relating precipitation and gamma-ray count rate. We demonstrate the ability of a contextually-aware city-scale sensor network to monitor moment-to-moment shifts in surface-level radioisotope concentrations. Additionally, we document the successful application of non-negative matrix factorization (NMF) to train static- and radon-enhanced background models according to a spectral triage method, as presented in Ref.~\cite{Bandstra2023}.

\begin{figure}[!t]
\centering{\includegraphics[width=3.5in]{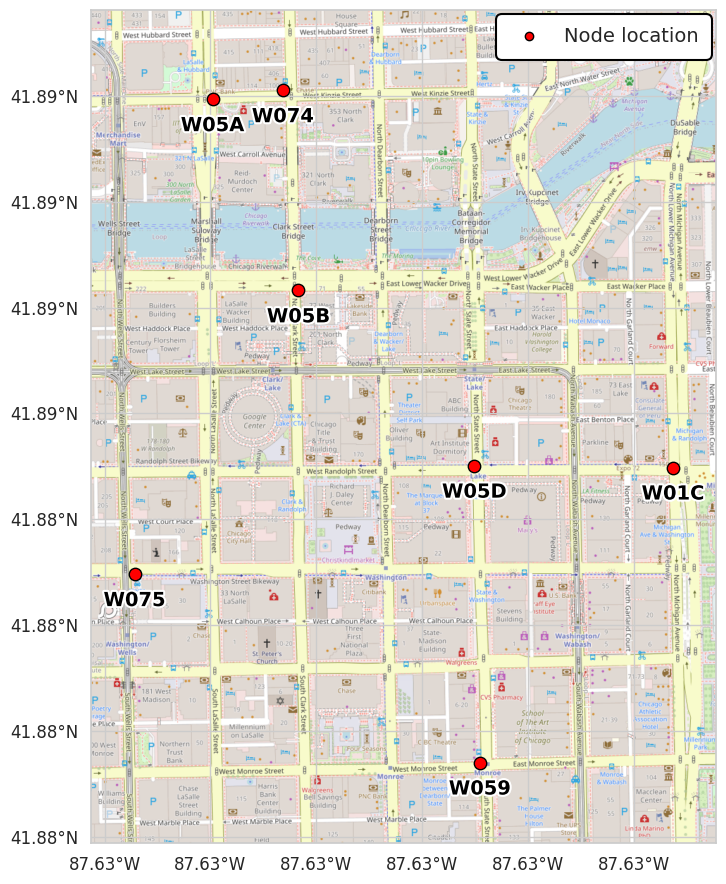}}
\caption{The locations of the selection of PANDA-DAWN nodes used in this analysis. These nodes were selected due to their proximity to one another and the completeness of their data over the time frame we consider.}
\label{fig:nodelocations}
\end{figure}

\begin{figure}[!t]
\centering{\includegraphics[width=3.5in]{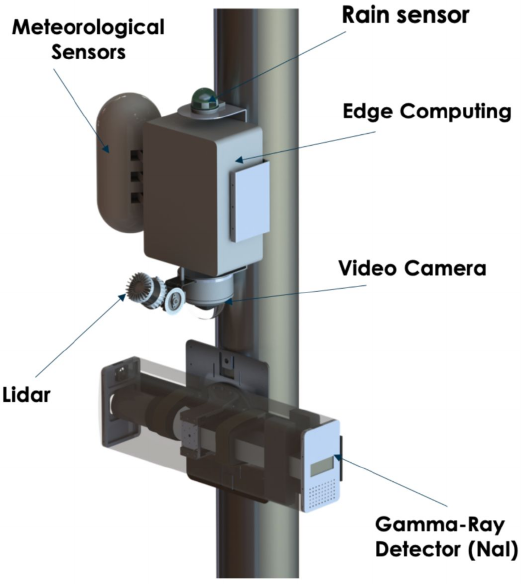}}
\caption{A diagram of a PANDA-DAWN sensor node and its most notable components. Used with permission from Ref.~\cite{Bandstra2023}.}
\label{fig:PandaNode}
\end{figure}

\section{Methodology}
\label{sec:methods}

We use data from a network of PANDA-DAWN sensor systems centered on ``The Loop'' neighborhood of Downtown Chicago, as depicted in Fig.~\ref{fig:nodelocations}. Our analysis focuses on a six-month period from September 2023 through February 2024.

\subsection{Instrumentation}
\label{subsec:instrumentation}

As depicted in Fig.~\ref{fig:PandaNode}, each DAWN sensor node contains a $2''\!\times4''\!\times16''$ NaI(Tl) radiation detector, which measures the energy and timing of incoming gamma rays. The unit features a solar awning and temperature regulation system to maintain performance in a wide range of conditions. A Hydreon RG-15 solid-state tipping bucket rain gauge estimates precipitation levels, while a BME680 sensor tracks the exterior temperature, pressure, and relative humidity. Additionally, each node is equipped with a VK-162 GPS receiver, and XNV-8081Z Hanwha outdoor network dome camera. The DAWN nodes leverage the Waggle platform~\cite{Waggle, waggle_sensor} that facilitates the deployment of powerful real-time edge computing algorithms by intelligently distributing applications over the node's computing resources: 2 NVIDIA Jetson Xavier NX boards (6 cores/8GB RAM, 1 GPU), and 2 RPis (4 cores/8GB RAM). The DAWN platform also leverages the Waggle data storage and dissemination tools further developed and extended by Sage, a National Science Foundation-funded cyberinfrastructure project designed to handle large sensor datasets for cutting-edge artificial intelligence and machine learning initiatives~\cite{sagecontinuum}.

The inclusion of sensors monitoring meteorological variables other than precipitation (\textit{i.e.}, temperature, pressure, and relative humidity) is well motivated by a wealth of literature describing their relationships with radon flux and radon daughter atmospheric concentration, both of which impact the ambient gamma radiation levels during non-rainy periods with weak winds~\cite{Szegvary2007, Shimo2007, Blaauboer1997, Minato1980, Takeuchi1982}. Pressure differentials are an important mechanism for radon gas exchange at the soil-air interface~\cite{Wilkening1981}, and pressure variation has been associated with large increases in the rate of radon exhalation~\cite{Kraner1964, Clements1974}. Long-term fluctuations in atmospheric $^{222}$Rn have been found to be positively correlated with temperature, while short-term variations are inversely correlated~\cite{Gaso1994}.  High humidity levels also result in increased water droplet condensation in the lower atmosphere, trapping local atmospheric radon progeny~\cite{Blaauboer1997}.

Tracking meteorological variables helps recognize the development of inversion layers, in which temperature rises with height until it declines beyond a few meters of altitude. Temperature inversions cause significant increases in the surface-level concentration of $^{222}$Rn and $^{222}$Rn progeny~\cite{Blaauboer1997, Israel1965, Melintescu2018, Yakoleva2016, Fujinami1995, Chambers2009}. Inversion layers form on calm nights and reduce vertical mixing, confining airborne radon near the surface. After sunrise, the lower atmosphere warms and the inversion layer usually dissipates. Diurnal cycles of radon concentration are stronger in dry conditions with minimal wind~\cite{Barbosa2017, Zahorowski2008} and occur more frequently in winter~\cite{Liu1984}. Inversion layers can sometimes lead to increased ambient gamma radiation levels comparable to precipitation-induced spikes~\cite{Barbosa2017}. Background estimation schemes that alternate between static- and radon-enhanced models must account for such situations which lack precipitation as an obvious indicator of radon enhancement.

The PANDA-DAWN nodes record temperature, pressure, and relative humidity data at $\frac{1}{30}$\,Hz, precipitation data at 1\,Hz. Each node is also capable of capturing and processing camera data at 25 frames per second. List-mode radiation data are integrated into 30-second spectra, providing the temporal granularity to detect moment-to-moment changes in the radiological conditions while suppressing Poisson noise. The energy bounds of the recorded spectra are 50\,keV to 3000\,keV to minimize low-energy background noise. To improve statistics in high-energy bins, we set the spectral bin width to be proportional to the square root of the incoming gamma-ray's energy.

\subsection{Analysis}
\label{subsec:analysis}

We employ the Berkeley Anomaly Detection (BAD) software to perform non-negative matrix factorization (NMF) on time-series spectra~\cite{BAD}. Non-negative matrix factorization is a dimensionality-reduction technique that extracts components making up the given training spectra~\cite{Bandstra2020, Lee1999}. The components can be associated with various physical effects and maximum-likelihood weights are assigned to each component for any given spectrum. After training, the weights can be fit to incoming spectra in real-time. For instance, creating two NMF components from spectra measured outdoors in varying weather conditions has empirically produced one component related to the static background (\textit{e.g.}, from cosmic rays and naturally occurring radioactive materials such as potassium, uranium, and thorium isotopes) and one component related to the $^{222}$Rn-dependent background~\cite{Bandstra2023}. Maximizing the mean count rate and minimizing the variance of one component (which becomes the static component) results in an even cleaner separation between the static and radon-dependent background~\cite{Bandstra2023}. Training three components further splits the radon progeny-dependent component into $^{214}$Pb- and $^{214}$Bi-associated components~\cite{Bandstra2023}. Non-negative matrix factorization is well suited for this problem because gamma ray counts sum linearly and are inherently discrete and positive~\cite{Bandstra2023}. Non-negative matrix factorization enables high-performance anomaly detection and isotope identification, and significantly outperforms other more conventional algorithms~\cite{Bilton2019}.

We designate periods of time as ``static,'' ``radon-enhanced,'' or ``anomalous'' according to the spectral triage methodology detailed in Ref.~\cite{Bandstra2023}. To classify radon-enhanced periods, we calculate the radon progeny proxy (RPP), developed in Ref.\cite{Bandstra2023}, based on the measured precipitation levels for the full measurement time. Radon progeny proxy grows proportionally with precipitation rate but decays with the half-life of $^{214}$Pb, the longer-lived of the two most relevant $^{222}$Rn daughters. The ``radon-enhanced'' label is applied to any period where the RPP is greater than a pre-determined threshold of 0.05. Static periods are fit to one NMF component, while radon-enhanced periods are fit to two. The training length is based on our desired false alarm rate. We train bootstrapped components over the first 5\,hours of ``static'' data to achieve a $\frac{1}{5}$\,hr$^{-1}$ false alarm rate. Similarly, two-component radon model is bootstrapped from the first 5\,hours of `radon-enhanced' data. The final models we use to analyze all subsequent data incorporate 80\,hours of ``static'' and 40\,hours of `radon-enhanced' data, which respectively result in false alarm rates of $\frac{1}{80}$\,hr$^{-1}$ and $\frac{1}{40}$\,hr$^{-1}$ for the static and radon-enhanced models.

To identify anomalous spectra, we use BAD to calculate an alarm metric that indicates the presence of a given isotope in the spectrum. BAD calculates the alarm metric by comparing two hypotheses: background-only versus background-plus-source. The metric is the difference in log-likelihoods between these fits, quantifying how much the inclusion of the source isotope improves the model's ability to explain the measured spectrum. We consider 24 potential sources, using energy emission distributions developed for the RADAI and REX projects~\cite{peplow_threat_2024}. These sources include naturally occurring radioactive material, isotopes commonly used in medical and industrial applications, and nuclear material. If the alarm metric crosses a threshold defined by our desired false positive rate, the anomaly is labeled with the isotope corresponding to the greatest alarm metric. Closely spaced alarms are clustered into continuous source encounters. For this analysis, we filter out non-$^{222}$Rn encounters to ensure changes in ambient count rate are due to environmental conditions. Additionally, we extract NMF components and weights from the presented data set to understand how each component develops over time and reacts to changing conditions.

\section{Results}
\label{sec:results}

\begin{figure}[!t]
\centering{\includegraphics[width=3.5in]{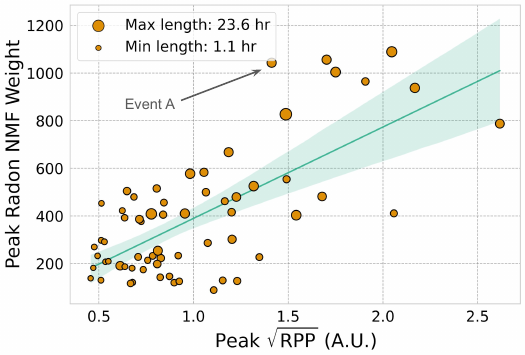}}
\caption{A regression between peak radon NMF weight and peak radon progeny proxy on an event-wise basis from Sept. 1\textsuperscript{st} 2023--March 1\textsuperscript{st}, 2024. The 95\% confidence interval on the regression line is shaded in green. The area of the marker is proportional to the square root of the length of the event. Event A is marked because it is a notable outlier.}
\label{fig:Regression}
\end{figure}

While understanding the relationship between precipitation and ambient count rate is not the primary focus of this work, we explore these variables to obtain a high-level perspective on the distribution of interesting precipitation events in our dataset. A precipitation event is defined as a continuous ``radon-enhanced'' period as defined in Sec.~\ref{subsec:analysis}, of at least 15\,minutes. A precipitation event is considered finished when its RPP falls below a pre-determined threshold of 0.2 for longer than an hour. Defined like this, most precipitation events in the dataset are between 2 and 5 hours long. In Fig.~\ref{fig:Regression}, we conduct a least-squares regression ($R^2 = 0.452$) between the peak radon NMF weight and the square root of the peak RPP of precipitation events from September 2023 through February 2024. We plot the peak values as the mean values can obscure notable features of events with long periods of unremarkable activity. Due to differing environmental geometry, each node has a unique static baseline reading and response to precipitation/source encounters. Placing nodes in open areas with an unobstructed sky view would minimize structure-related shielding and thus reduce the node-to-node spread in baseline count rates. However, that advantage must be weighed against the competing goal of positioning detectors where they intercept the greatest vehicle traffic and increase the monitored area. To remove unavoidable site-specific offsets, one can normalize each detector’s data to its own long-term mean static rate before events are compared across the network, as we do in the analysis in Fig.~\ref{fig:IRRCTimeSeries} by applying a node-by-node linear-regression correction. For consistency, Fig.~\ref{fig:Regression} displays data collected from a single node located at the intersection of North Michigan Avenue and East Randolph Street, Chicago. We find a moderate trend between the peak NMF radon weight and the square root of the peak RPP. This finding aligns with several previous studies~\cite{Greenfield2003, Mercier2009, Bottardi2020}, which identify power-law relationships with similar coefficients between ambient gamma radiation levels and precipitation. Additionally, it is physically consistent with findings that demonstrate the vast majority of scavenged radon progeny can be deposited by minimal amounts of rain~\cite{Barbosa2017, Paatero2000, Yoshioka1992}. Further rain leads to progressively smaller contributions to surface-level radiation and dilutes the radon progeny, causing a net reduction in radon progeny concentration in rainwater. Furthermore, longer rain events deposit more radon progeny and increase count rate, an effect noted in~\cite{Fujitaka1992}.

While most precipitation events follow the described trend, we find some notable outliers. Event A (as labeled in Fig.~\ref{fig:Regression}) has a significantly greater peak NMF weight than predicted from its peak RPP. Since the RPP captures the effects of amount of precipitation, and local shielding does not change over time, the most likely explanation to the deviation is in the radon progeny concentration in the rainwater. The most frequent cause of large variations in radon progeny concentration is the origin of the air mass, as oceans do not release significant amounts of $^{222}$Rn. The heightened radon concentration of air masses that have recently passed over land is well-described in existing literature~\cite{Arnold2010, Barreira1961, Bottardi2020, Burnett2010, Inomata2007, Israel1965, Liu1984, Paatero2000, Wilkening1981}. Paatero \textit{et al.} find variations in radon progeny concentration in precipitation spanning four orders of magnitude due to warm/cold fronts and air mass origin~\cite{Paatero2000}.
    
Other similarly anomalous precipitation events have been explained through exceptional meteorological phenomena. Inomata \textit{et al.} attribute a particularly large spike in gamma radiation levels to the coincident arrival of a radon-rich continental air mass, near-surface air mass convergence, and strong precipitation~\cite{Inomata2007}. Barbosa \textit{et al.} ascribe the abnormally low radon progeny deposition of a precipitation event to its cumuliform clouds that developed under conditions where the air was cooler than the ground, causing rapid accretion of liquid water and dilution of the radon progeny. Our goal is to characterize these extraordinary events using basic, easily-measured meteorological variables to inform background models of anomaly detection algorithms in the same class as that presented in Sec.~\ref{subsec:analysis}. 

\subsection{Event A: Case study of a precipitation event indicating anomalously large radon deposition}

\begin{figure}[!t]
\centering
\includegraphics[width=3.5in]{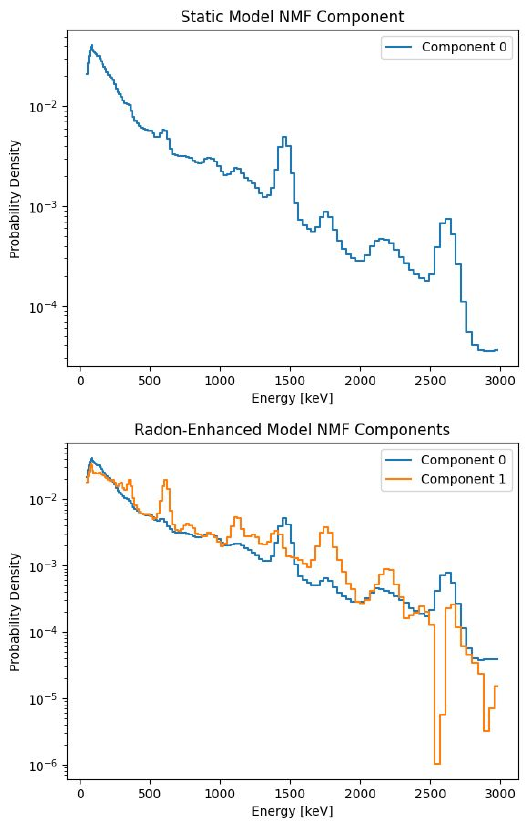} 
\caption{NMF components for the static model (upper) and the radon-enhanced model (lower).}
\label{fig:Components}
\end{figure}

Ultimately, a general characterization of anomalous precipitation events is desired. However, it is informative to perform a detailed analysis on a singular case to demonstrate the kinds of information that can be gleaned from the approach described in Sec.~\ref{sec:methods}. In this section, we examine Event A from Fig.~\ref{fig:Regression}, a precipitation event with a much greater peak radon NMF weight than predicted based on its peak radon progeny. This event took place in the early morning of December 1\textsuperscript{st}. We pull this data from the same node as used in Fig.~\ref{fig:Regression}.

\begin{figure}[!t]
\centering
\includegraphics[width=3.5in]{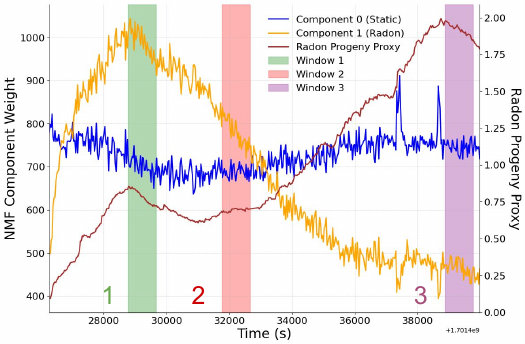} 
\caption{A time series of the radon progeny proxy (purple), static-associated NMF component weight (blue), and radon-associated NMF component weight (orange) throughout event A. Three averaging windows (1, 2 and 3) correspond to the spectra in Fig.~\ref{fig:Overlapped}.}
\label{fig:RainMetrics}
\end{figure}

The NMF components for both the static and radon-enhanced models are displayed in Fig.~\ref{fig:Components}. Figure~\ref{fig:RainMetrics} shows a time series of the radon progeny proxy and NMF component weights (for the radon-enhanced model) during Event A. We observe the typical shape of the ambient gamma radiation level during precipitation events to be, as described by Ref.~\cite{Yakoleva2016}, a sharp linear increase with a bell-shaped peak followed by a quasi-exponential decay. The radon progeny proxy initially closely tracks the radon-associated NMF component weight in shape, until it diverges as precipitation continues to fall but the radon progeny content of the cloud is depleted. One potential explanation for a difference in decay rate between RPP and the radon component weight is local topography influencing the collection and dispersion of runoff. This has been posited as an explanation for abnormally low gamma radiation levels during certain precipitation events by Ref.~\cite{Takeyasu1993}. 

\begin{figure*}[!t]
\centering
\includegraphics[width=\textwidth]{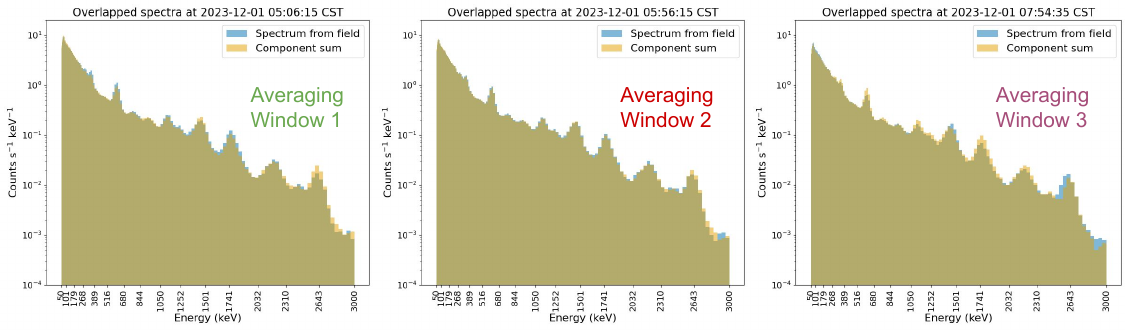}
\caption{The 15-minute averaged field spectrum (pale blue) and spectrum constructed from the inner product of the NMF components and weights (pale yellow). The averaging windows correspond to those highlighted in Fig.~\ref{fig:RainMetrics}}
\label{fig:Overlapped}
\end{figure*}

Fig.~\ref{fig:Overlapped} overlays the weighted sum of the NMF components with the incoming spectrum, both continuously averaged over 15~minutes throughout Event A. We observe that the NMF components closely model the field spectrum during the radiological peak. However, the NMF component sum loses accuracy as the count rate falls and the event enters a more ordinary regime, which causes undesirable $^{226}$Ra alarms. This indicates that specificity may be improved by training models for differing meteorological conditions. 

\begin{figure}[!t]
\centering
\includegraphics[width=3.5in]{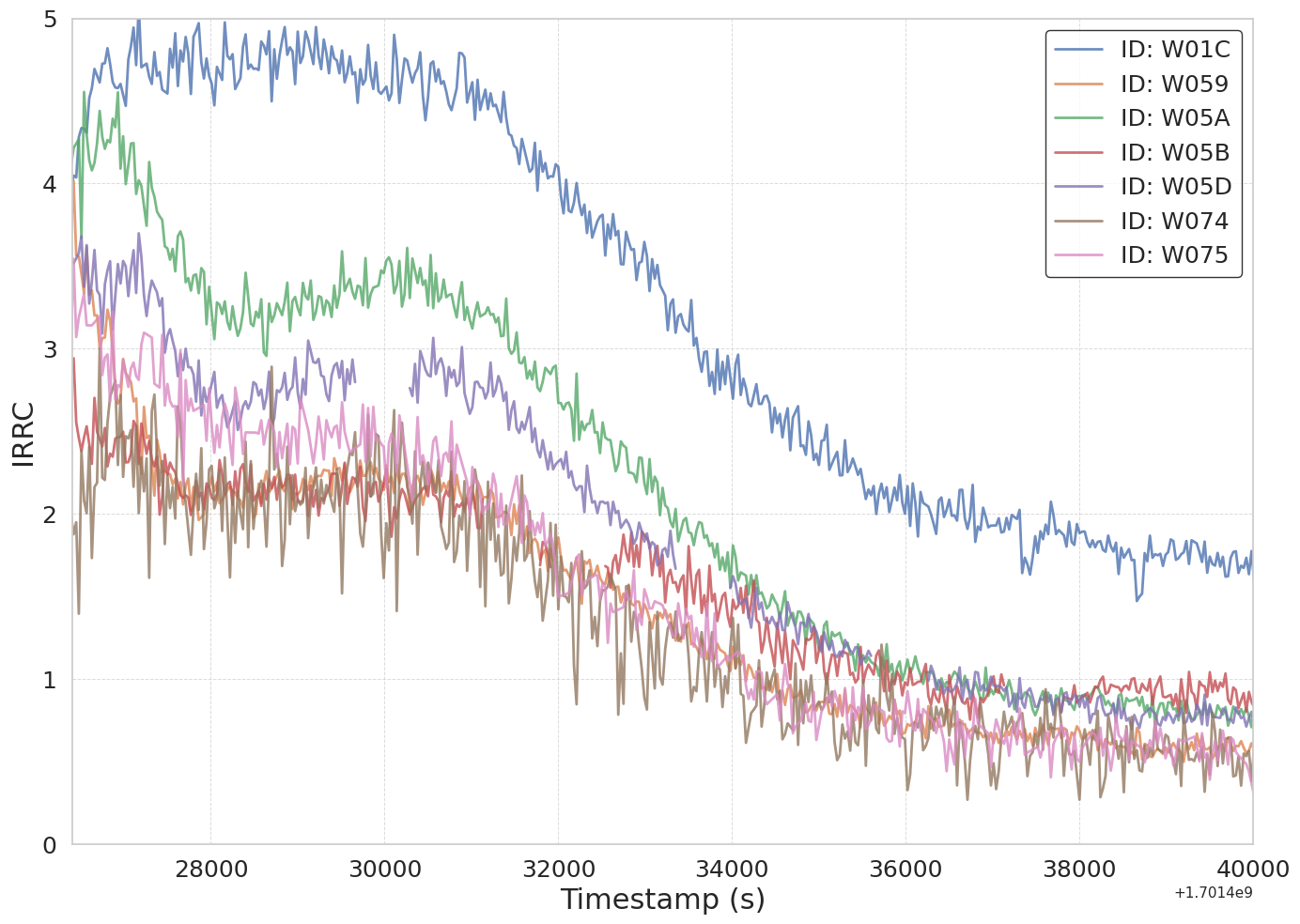}
\caption{A time series of IRRC at each node during Event A. Minor gaps are due to non-$^{226}$Ra alarms.}
\label{fig:IRRCTimeSeries}
\end{figure}

Since we interpret differences between the measured and rain-predicted ambient radon levels as variations in rain radon concentration, we define the implied relative radon concentration (IRRC) as the instantaneous ratio between the measured and predicted radon NMF weight. The predicted radon NMF weight is obtained through a linear regression between RPP and radon NMF weight over the entire dataset for a given node. Fig.~\ref{fig:IRRCTimeSeries} displays the evolution of IRRC over the course of Event A in all nearby sensors. The regression used to calculate IRRC was created independently for each sensor using data from September 1st through the conclusion of Event A. We observe that the onset of heavy precipitation produces a regional spike in IRRC, which gradually falls. The easternmost sensors have the latest peaks, most prominent in sensor W01C. This indicates that the spatial distribution of the IRRC moves east over time.

\begin{figure}[!t]
\centering{\includegraphics[width=3.5in]{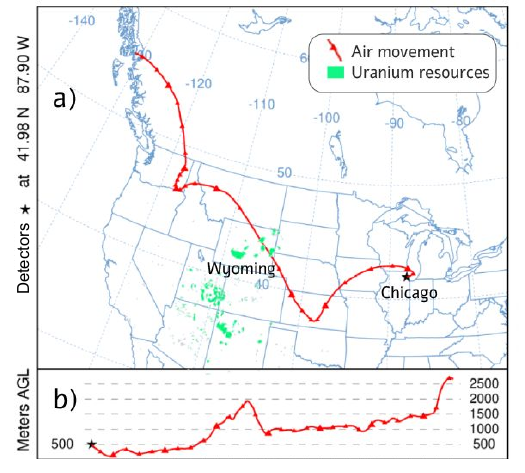}}
\caption{In the upper plot (a), the path of an air mass over the 7 days prior to depositing an anomalously large amount of radon progeny in Chicago. In the lower plot (b), the height of the air mass (in meters) through its history. Both plots were created using the NOAA's HYSPLIT model. In green, major uranium deposits in the United States, which are extracted from a public domain image provided by the United States Geological Survey~\cite{usgs_uranium_2022}.}
\label{fig:Trajectory}
\end{figure}

Fig.~\ref{fig:Trajectory} shows the air mass's history, which we estimate using the National Oceanic and Atmospheric Administration's publicly-available HYSPLIT model—a standard tool for back-trajectory analysis in radon-related meteorological studies~\cite{Chambers2009, Inomata2007, Mercier2009}. The recent history of the air mass is purely continental, where we expect the greatest radon exhalation. The air mass drifted over the uranium-rich regions of northeast Wyoming mere days before the precipitation event in Chicago. Of the lowest IRRC events, none approached any major uranium deposits. We hypothesize that the major influence on the radon progeny concentration of this event is the trajectory of the associated air mass. This observation aligns with Barreira \textit{et al.}, who also observed heightened radon concentration from air masses passing over uranium-rich regions~\cite{Barreira1961}.

As precipitation continues, the IRRC of the event gradually diminishes as the air mass's radon progeny is depleted. Towards the end of the event, the IRRC falls such that less radon progeny is being detected than expected, even for the relatively low amount of rainfall. Besides the atmosphere's radon progeny content being further exhausted, several factors could contribute to this phenomenon. Previous authors have suggested that water-logged soil attenuates gamma rays from radon daughter decays within the soil~\cite{Gaso1994, Malakhov1966}. Additionally, as water fills the pores of the soil, less radon is exhausted, causing a reduction in the indirect radon contribution~\cite{Szegvary2007}. Furthermore, surface runoff is always a factor, as radon progeny can be removed to other locations depending on the local topography, especially in urban environments which lack porous ground surfaces. 

We emphasize the importance of taking measurements from a sensor network in this analysis, as opposed to using an individual sensor. If readings from only a single sensor were available, differences between the actual and predicted radon concentration could potentially be explained by local topography, temporary shielding, sensor malfunction, or other unexpected hyper-local phenomena. If the network was too sparse (\textit{e.g.},\textless{}1 detector /10000\, km$^2$), it would be difficult to ascribe correlated changes between nodes to the same local (km-scale) phenomena. Because the evolution of the IRRC is consistent throughout the area, we can reliably attribute the unique characteristics of the radiological readings to the properties of the meteorological event.

\subsection{Characteristics of implied relative radon concentration}
\label{subsec:general}

To optimize the performance of anomaly detection, PANDA is developing adaptive algorithms that can autonomously alternate between background models in response to real-time changes in environmental conditions. The shape of the radon NMF component plausibly depends on the type of meteorological event (\textit{e.g.}, in terms of peak height). Understanding the characteristics of radon-enriched compared to radon-depleted precipitation events can potentially enable algorithms to switch background models in real-time, enhancing detection efficiency. 

\begin{figure}[!t]
\centering{\includegraphics[width=3.5in]{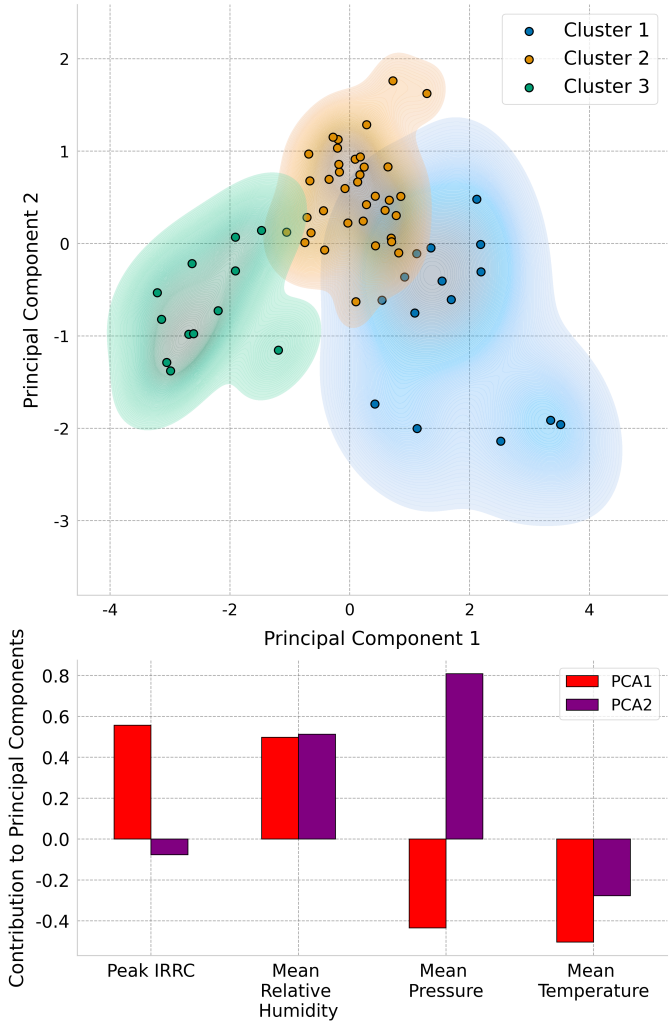}}
\caption{In the upper plot (a), clustering precipitation events in our stated time period based on a principal component decomposition of meteorological variables and peak implied relative radon concentration. The shading around eligible clusters is created with kernel density estimations. In the lower plot (b), the composition of each principal component. Event A has the greatest value of PCA2.}
\label{fig:PCA}
\end{figure}

To characterize radon-enriched precipitation events, we perform a principal component analysis on event-wise data, using peak IRRC, mean temperature, mean humidity and mean pressure as predictors. In Fig.~\ref{fig:PCA}, we group the events based on the first two components using a $k$-means clustering algorithm and summarize the composition of each component. The algorithm produces three distinct clusters. Cluster 1 (blue) roughly captures humid, low-pressure systems, typically with long-lived rainbands that efficiently scavenge radon and drive the highest IRRC peaks. Cluster 2 (orange) appears to correspond to light stratiform rain, characterized by slightly above-average pressure and humid but shallow boundary layers. Cluster 3 (green) represents weak-gradient or lake-breeze convective showers with, lower humidity, warmer boundary-layer temperatures, and brief, spotty precipitation that yields the weakest radon wash-in.  

\section{Discussion}
\label{sec:discussion}
The thrust of this work is that there exist opportunities to improve adaptive radioactive anomaly detection algorithms using easily-measured meteorological variables collected by networks of sensor systems. One option is to train several radon-enhanced background models using the meteorological variables associated with the events as features. For instance, a different model may be trained for each of the clusters in Fig.~\ref{fig:PCA}. This logic is similar to that used by~\cite{Bandstra2023}, which suggests differing background models due to direct and indirect contributions. Just as our anomaly detection algorithm uses different characteristic spectra for varying levels of uranium enrichment, differing NMF components can be trained for clouds of varying levels of radon enrichment. For any incoming precipitation event, the algorithm can begin with a prior probability for the type of meteorological event and perform Bayesian updates based on real-time meteorological data. From these probabilities, the algorithm can decide which background model to use. One important limitation to this concept is that including more background models necessitates the collection of significantly more data during training, which would increase the delay from system activation to anomaly detection in exchange for an uncertain improvement in performance. These trade-offs should be rigorously evaluated in future studies.

We have shown that an extreme case of precipitation-induced radon enhancement was potentially due to its path intercepting a region of heightened radon exhalation in Wyoming. This could indicate that collecting that air mass path information is useful for understanding the expected radiological characteristics of a precipitation event. Going forward, we emphasize that sensor systems, such as those in the PANDA-DAWN network, should consider air mass paths when creating background models. This will reduce the uncertainty in the expected radon contribution, allowing sensors to become more sensitive to radiological anomalies or threats. If more frequent indications of air mass origin are desired, we suggest monitoring instantaneous wind direction, as previous studies have successfully used local two-dimensional wind direction as a proxy for air mass origin~\cite{Arnold2010, Zahorowski2008}.

One limitation to our analysis method is that we label alarms according to the isotope that produces the greatest alarm metric, as defined in Sec.~\ref{subsec:analysis}. The metrics for different isotopes sample different distributions (with different thresholds) and are not directly comparable on an absolute scale. A more robust labeling method would account for the variance from the Fischer information matrix associated to these metrics, and a similarity matrix between isotopes. While we do not expect this issue to impact the presented results, an improvement in this area would improve the specificity of our source identification in future studies. 

In future work, we hope to take advantage of the camera data provided at each node. By angling the camera upwards, one could potentially detect the size, shape and height of visible clouds and use this information to determine the expected level of radon exposure. For this, we can also leverage existing Sage plugins that find cloud cover, type and movement~\cite{weather_classification_2025, cloud_cover_2025, cloud_motion_2024}. The camera footage can also provide information regarding changes in traffic density through the use of object recognition algorithms such as YOLOv7~\cite{YOLOv7}. An increase in the number of cars on the road could potentially shield the detectors from gamma rays originating in soil and concrete.

While this article narrowly focuses on characterizing radon progeny deposition events, this work also demonstrates the potential of sensor arrays in urban environments in radiation measurement and anomaly detection. As depicted in Fig.~\ref{fig:IRRCTimeSeries}, we successfully used a sensor to model the moment-to-moment non-static radioactivity. By interpolating values between nodes, this method of analysis can be used for source localization, air mass tracking, and instantaneous radioactive fallout mapping. In the future, we hope to leverage inter-node communication to produce more precise evaluations of environmental radiation and more accurate anomaly detection techniques.

\section{Conclusions}
\label{sec:conclusions}

We have conducted an investigation of the meteorological characteristics of radon-enhanced wet deposition phenomena, using data collected from a sensor array in Chicago, IL. Our case study of a precipitation event particularly rich in radon progeny highlighted the importance of air mass history in anticipating the radon content of radon deposition events. We observed that non-negative matrix factorization components may differ between typical and anomalous radon-enhanced events, suggestive of potential sensitivity gains from training multiple background models to encapsulate all scenarios. Our unique contextually-aware sensor network allowed us to monitor the spatial and temporal evolution of the implied radon progeny concentration. Additionally, we studied the distribution of implied radon concentration as it varies with season and other meteorological characteristics. We hope to improve upon this work by making use of the visual camera data obtained by each sensor node. Future work should focus on quantifying the improvement provided by multiple background models and balancing those gains against the challenges of acquiring sufficient training data in a timely manner. Our work demonstrates the potential of sensor arrays in urban environments for detailed monitoring and analysis of radon progeny deposition events, contributing to the development of more precise and responsive radiological threat detection algorithms.

\bibliographystyle{IEEEtran}
\bibliography{references}

\end{document}